**Title:**

**A generalized linear model for decomposing *cis*-regulatory, parent-of-origin, and maternal effects on allele-specific gene expression**

Yasuaki Takada[1], Ryutaro Miyagi[2], Aya Takahashi[2,3], Toshinori Endo[1], Naoki Osada[1*]

1 Graduate School of Information Science and Technology, Hokkaido University

2 Department of Biological Sciences, Tokyo Metropolitan University, Hachioji, Japan

3 Research Center for Genomics and Bioinformatics, Tokyo Metropolitan University, Hachioji, Japan

Corresponding Author: Naoki Osada, nosada@ist.hokudai.ac.jp

Graduate School of Information Science and Technology, Hokkaido University

Kita14-Nishi9, Kita-ku, Sapporo, Hokkaido 060-0814, Japan

Keywords: gene expression, *cis*-regulatory mutation, genomic imprinting, maternal effect

Running title: Decomposition of the effects on gene expression

**ABSTRACT**

Joint quantification of genetic and epigenetic effects on gene expression is important for understanding the establishment of complex gene regulation systems in living organisms. In particular, genomic imprinting and maternal effects play important roles in the developmental process of mammals and flowering plants. However, the influence of these effects on gene expression are difficult to quantify because they act simultaneously with *cis*-regulatory mutations. Here we propose a simple method to decompose *cis*-regulatory (*i.e.*, allelic genotype, AG), genomic imprinting (*i.e.*, parent-of-origin, PO), and maternal (*i.e.*, maternal genotype, MG) effects on allele-specific gene expression using RNA-seq data obtained from reciprocal crosses. We evaluated the efficiency of method using a simulated dataset and applied the method to whole-body *Drosophila* and mouse trophoblast stem cell (TSC) and liver RNA-seq data. Consistent with previous studies, we found little evidence of PO and MG effects in adult *Drosophila* samples. In contrast, we identified dozens and hundreds of mouse genes with significant PO and MG effects, respectively. Interestingly, a similar number of genes with significant PO effect were detect in mouse TSCs and livers, whereas more genes with significant MG effect were observed in livers. Further application of this method will clarify how these three effects influence gene expression levels in different tissues and developmental stages, and provide novel insight into the evolution of gene expression regulation.

**INTRODUCTION**

Epigenetics, which refers to phenotypic modifications in the absence of changes to information encoded in DNA molecules, has become a central topic in biological research in order to understand the development of multicellular organisms and maintenance of highly differentiated cells and tissues (WADDINGTON 1942). Although epigenetic effects can contribute to a wide array of phenotypes, most studies of epigenetic effects in the era of molecular biology have concerned gene expression, which is much more easily quantified than other phenotypes on a genome-wide scale. Epigenetic effects on gene expression can be classified as either *cis-* or *trans*-epigenetic effects (BONASIO *et al.* 2010). In diploid organisms, *cis*-epigenetics refers to chromosome-specific modification of gene expression. For example, histone protein modification and cytosine methylation could affect the expression of genes located on the same chromosome. In contrast, *trans*-epigenetics refers to epigenetic modifications of gene expression that have equal effects on both chromosomes of diploid organisms. In a broad sense, *trans*-epigenetics would therefore include all gene expression changes caused by intrinsic and extrinsic environmental changes, such as those observed in cell differentiation and reaction to environmental change.

Genomic imprinting is a well-known phenomenon in mammals and flowering plants and refers to the process by which genes inherited from a particular sex are down-regulated or completely silenced (KÖHLER *et al.* 2012; BARLOW AND BARTOLOMEI 2014). By the above definition, genomic imprinting is caused by *cis*-epigenetic mechanisms. Among mammals, genomic imprinting has been most extensively studied in laboratory mice (*Mus musculus*), and approximately 150 loci, including both protein-coding genes and non-coding RNAs, have been experimentally identified as imprinted (BLAKE *et al.* 2010). In contrast, it remains unclear whether genomic imprinting can be detected in non-mammalian animals. In particular, there have been conflicting results whether fruit flies (*Drosophila melanogaster*), which lack DNA methyltransferases except for the *Dnmt2* (*MT2*) product, are subject to genome-wide imprinting effect (MENON AND MELLER 2010; COOLON *et al.* 2012; MCEACHERN *et al.* 2014; TAKAYAMA *et al.* 2014). Although the underlying mechanisms and causes of imprinting are not entirely clear, genomic imprinting is necessary to our understanding of the complex relationships between genotypes and phenotypes (FERGUSON-SMITH 2011). Therefore, the effects of genomic imprinting in different organisms should be determined using standardized methods.

Recent advance in sequencing technology has enabled the evaluation of genome-wide imprinting pattern. RNA-seq transcriptome sequencing has allowed the measurement of chromosome-specific (or allele-specific) gene expression levels for paternally and maternally inherited genes that harbor genetic markers such as single nucleotide variation (SNV)(WITTKOPP 2005). Comparison of patterns of allele-specific gene expression between reciprocal crosses is informative because of potential differences in gene expression levels consequent to *cis*-regulatory mutations (WITTKOPP 2005); *i.e.*, observation of parent-of-origin (PO)-dependent allelic imbalance in both reciprocally-crossed individuals suggests genomic imprinting rather than a *cis*-regulatory effect. Accordingly, such comparisons are widely used to discern *cis*-genetic and *cis*-epigenetic effects; *i.e.*, if allelic imbalance depending on PO is observed in both reciprocally-crossed individuals, the imbalance is likely due to genomic imprinting rather than the *cis*-regulatory effect. Several studies have implemented these strategies to identify genes subject to genomic imprinting on a genome-wide scale (BABAK *et al.* 2008; GREGG *et al.* 2010; COOLON *et al.* 2012; CALABRESE *et al.* 2015). However, this method tends to be conservative if the *cis*-regulatory effect is prevalent, because it may reduce the power of statistical tests to detect imprinting effects.

In addition, comparisons of reciprocal crosses are complicated by additional confounding factors because reciprocally crossed individuals have different maternal environments. This finding was first described as the maternal effect in a classical experiment by Walton and Hammond (WALTON AND HAMMOND 1938). Although classical family studies and embryo transplantation studies have shown that the environmental effect on offspring phenotype is generally larger than the genetic effect (GLUCKMAN AND HANSON 2004), genetic effects may contribute to the maternal effect to some extent. One example would be the genotype effect in oocyte cytoplasm, as these cells inherit mRNA and mitochondrial DNA from the mother in a process that meets the definition of a *trans*-regulatory effect, which is a genetic effect equally affecting to both chromosomes by diffusible way (EMERSON AND LI 2010). In addition, maternal genotype (MG)-determined prenatal and postnatal environments will contribute to offspring phenotypes. Here we use the term MG effect, assuming appropriate control of non-genetic environmental factors. Although the MG effect may be subtle, it might contribute to gene expression pattern in a *trans* manner. The PO and MG effects are hardly distinguished in a conventional genetic analysis, because the phenotypes of offspring are defined by the sum of the maternally- and paternally-inherited alleles and the maternal and paternal contributions are unseparatable in phenotype (HAGER *et al.* 2008). However, by directly measuring gene expression level of maternally- and paternally-inherited alleles, there is an oppotunity to separately evaluate the PO and MG effects. A previous study proposed a method to jointly estimate the genetic *cis*-regulatory, or allelic genotype (AG) and PO effects, but the MG effect has been neglected (ZOU *et al.* 2014).

Here, we have proposed a simple statistical framework for simultaneously and separately estimating the AG, PO, and MG effects on gene expression in reciprocally-crossed individuals when the allele specific gene expression level is provided, and have demonstrated the effectiveness of this method using a simulated dataset. We used a generalized linear model (GLM) to quantify each effect, assuming a lack of interaction. The previous genome-wide study of the PO effect, which was designed without replication, suggested the importance of biological replicates (COOLON *et al.* 2012). GLMs efficiently deal with the contributions of each factor and fluctuations among biological replicates. We applied this method to two different organisms, *Drosophila* and mice. For the former, we obtained a new adult female whole-body gene expression dataset using two pairs of reciprocal crosses: F1 hybrids of the *Drosophila* Genetic Reference Panel (DGRP) strains for which genomic sequences were made publicly available (MACKAY *et al.* 2012). For mice, we reanalyzed recently published datasets of trophoblast stem cells (TSCs) and livers from reciprocal crosses between CAST/EiJ and C57BL/6J (Cast/B6) animals (GONCALVES *et al.* 2012; CALABRESE *et al.* 2015). Although we identified statistically significant AG effects for a considerable number of genes in both organisms, we found very small number of genes with significant PO and MG effects in *Drosophila*, consistent with an earlier report by Coolon (COOLON *et al.* 2012). In contrast, we found that dozens of genes in mouse TSCs and livers were subject to significant PO effects. In addition, considerably higher number of genes in the mouse liver exhibited significant MG effect compared to genes in the mouse TSCs, indicating that the MG effect tends to be tissue- or developmental stage-specific.

**METHODS**

*GLM design*

Suppose that there are two different isogenic strains, A and B. Following to a general rule, A × B would denote F1 hybrids generated by a cross between females of strain A and males of strain B. When strains A and B exhibit sufficient genetic differences, we could measure allele-specific gene expression levels using RNA-seq for each reciprocal cross, A × B and B × A, with biological replications. The allele-specific expression value $E$ would then be defined using the following linear regression model expression.

$$E \sim \mathrm{AG} + \mathrm{PO} + \mathrm{MG} + \varepsilon \quad (1)$$

, where AG, PO, and MG represent the effect of allelic genotype, parent-of-origin, and maternal genotype, respectively. Here, we assumed each effect was a fixed effect and assigned binary codes to the effects. For AG, we assigned values of 0 and 1 to A and B, respectively. For PO, we assigned a value of 0 if the chromosome was inherited from the mother, and 1 if the chromosome was inherited from the father. For MG, we assigned a value of 0 to sample A × B (maternal genotype A) and 1 to sample B × A (maternal genotype B). The error term was estimated using biological replicates of samples. A schematic representation of this design is shown in Figure 1.

We propose two GLM models to utilize allele-specific gene expression level to estimate the AG, PO, and MG effects. The first model is a log-normal GLM. In a typical RNA-seq data analytical pipeline, gene expression levels are normalized by gene length and total read count, and represented as FPKM values, which can be assumed to exhibit a log-normal distribution (BENGTSSON *et al.* 2005). Therefore, by log-transforming allele-specific FPKM values, we could apply a Gaussian distribution to the distribution of response variable in the GLM. The second model is a negative binomial GLM. Since an actual RNA-seq data is a count data represented by the number of reads mapped on the transcript sequences, a negative binomial model has been widely adopted in many statistical packages such as EdgeR (MᴄCᴀRTHY *et al.* 2012) and DESeq (AɴᴅᴇRs ᴀɴᴅ HᴜʙᴇR 2010) for analyzing RNA-seq data. The log-normal and negative binomial GLM analyses were performed using the glm function and EdgeR libraries in the R statistical package (R Project for Statistical Computing, Vienna, Austria), respectively. The R script and expression data files used for the GLMs are provided as Supplementary Data 1.

*Computer simulation*

In computer simulations, we only considered the log-normal GLM. We assumed normally distributed allele-specific gene expression levels with the fixed additive effects of AG, PO, and MG. Following the design shown in Figure 1, we considered eight different cases for the presence and absence of fixed effects: no effect, AG, PO, MG, AG + PO, AG + MG, PO + MG, and AG + PO + MG. As the statistical detection power for each fixed effect was determined by the magnitude of the fixed effect size relative to biological/environmental/statistical fluctuations, we evaluated the methodologic power using the ratio of the fixed effect to the standard deviation of experimental noise, which was equivalent to Cohen's $d$ statistics. A larger $d$ indicated more power for effect detection.   For 2- and 5-times replicated experimental designs, we changed $d$ values from 1 to 5 with 0.1 intervals and calculated true positive rates.

For each simulated gene, we arbitrarily assigned a basal gene expression level and added random noise drawn from a standard normal distribution $N(0, 1)$. After adding errors, a fixed effect was added to the expression value. For example, when $d = 5$, we added 5 to the expression value when the binary code of samples (Figure 1) was 1 for each fixed effect. For each condition, 1250 genes were simulated (in total 10,000 genes for one replicate) with two or five replications, and the simulated dataset was analyzed using the above-described GLM method. The significance of each gene test was evaluated using the criterion of FDR = 0.05 (BENJAMINI AND HOCHBERG 1995).

*RNA-seq dataset*

In this study, we obtained new gene expression data of two pairs of reciprocal crosses of *D. melanogaster* from the DGRP (MACKAY *et al.* 2012; MASSOURAS *et al.* 2012), representing crosses between RAL-324 and RAL-852 and between RAL-799 and RAL-820. These strains were arbitrarily chosen from a list of DGRP strains. The flies were grown at 25ºC with a 12-h light-dark cycle and were fed standard corn-meal fly medium. F1 virgin females were collected within 8 hours of eclosion and maintained separately on the regular food media. After 4–7 days of the isolation, 100 flies per sample were flash frozen in liquid nitrogen and stored at −80 ºC. The whole-body total RNA was extracted using the TRIzol Plus RNA Purification Kit (Thermo Fisher Scientific, Waltham, MA, USA). The concentration of extracted total RNA was measured using a Nanodrop 2000c (Thermo Fisher Scientific) and quality was evaluated using a TapeStation (Agilent Technologies, Foster City, CA, USA). For RNA-seq, 250 ng of total RNA were used for library construction with the TruSeq Stranded mRNA Library Prep Kit (Illumina, San Diego, CA, USA). Samples were barcode-indexed and pooled for each sequencing lane. Raw read data were deposited into a public database under the Bioproject ID PRJDB5381. The accession number and index type of each library are provided in Table S1. Mouse TSC expression data were retrieved from the GEO database (https://www.ncbi.nlm.nih.gov/geo/)

under the accession number GSE63968, and mouse liver data were downloaded from ArrayExpress (https://www.ebi.ac.uk/arrayexpress/) using the accession number E-MTAB-1091.

*Estimation of allele-specific expression data*

We obtained genomic sequences of focal strains to estimate allele-specific gene expression levels. For *Drosophila*, we used the version dm3 reference genome sequence, and obtained a VCF file (freeze 2.0 call) containing the information about the SNVs in DGRP strains from the DGRP website (http://dgrp2.gnets.ncsu.edu/). Genome sequences of RAL324, RAL799, RAL820, and RAL852 were reconstructed using the FastaAlternateReferenceMaker command in GATK software (MCKENNA *et al.* 2010). A mouse reference genome sequence (GRCm38) and VCF files of CAST/EiJ and C57BL/6NJ strains were retrieved from the ENSEMBL database (http://ensembl.org/) and the Sanger Mouse Genomes Project website (http://www.sanger.ac.uk/science/data/mouse-genomes-project), respectively. The genome sequences of CAST/EiJ and C57BL/6NJ were reconstructed using the same procedure described for *Drosophila* data.

We used ASE-TIGER software, which is based on Bayesian inference, to estimate the allele-specific FPKM and number of allele-specific mapped reads (NARIAI *et al.* 2016). Briefly, RNA-seq reads were mapped on transcriptome sequences reconstructed from two parental genomes. Strain-specific *Drosophila* and mice transcriptome sequences were generated from reconstructed genome sequences using the annotation file for the build 5 *D. melanogaster* genome (downloaded from NCBI: https://www.ncbi.nlm.nih.gov/) and Mus_musculus.GRCm38.84.gtf for mice (downloaded from ENSEMBL), respectively. We used bowtie2 software to map RNA-seq reads, using the option of "–very sensitive" (LANGMEAD AND SALZBERG 2012). Because we could not accurately estimate the allele-specific expression levels of genes with small numbers of SNVs within genes, we filtered out transcripts with <3 SNVs in the exons. Because ASE-TIGER reported FPKM and the number of mapped reads for each transcript, those values were summed across isoforms to estimate the value at the gene level. For the log-normal model, weakly expressed genes (average FPKM <0.1) were filtered out. Before log-transformation, we replaced FPKM values <0.01 with 0.01 to avoid legalism associated with very small or 0 values. For the negative binomial GLM, genes with <1 CPM (count per million mapped reads) in less than half of the chromosomes were filtered out.

*Gene Ontology enrichment analysis*

We utilized the DAVID 6.7 webserver to identify significantly enriched Gene Ontology terms from a list of genes with significant effects (JIAO *et al.* 2012). Lists of background genes were extracted from all analyzed genes in each dataset. For each Gene Ontology term, terms with a $p$ <0.05, determined using a modified Fisher's exact test after correcting multiple testing, were selected as significantly overrepresented functional categories (HOSACK *et al.* 2003).

**RESULTS**

*Design of the GLM*

We conducted a GLM analysis in order to jointly estimate the effects of AG, PO, and MG. Two different GLMs, the log-normal and negative binomial GLM were applied. A full description of the GLMs is presented in the in the Materials and Methods section. Briefly, in the log-normal GLM, we estimated the allele-specific gene expression level as FPKM for each gene and transformed these values to a $\log_2$ scale. The $\log_2$-transformed expression values were used as response variables in the GLM assuming a Gaussian distribution. In the negative binomial GLM, the estimated number of reads mapped on the transcriptome sequences from each chromosome were used as a count data. Three fixed effects (AG, PO, and MG) were set as the explanatory variables in the model and binary codes were assigned to the values. A schematic representation of the model is shown in Figure 1.

*Computer simulations*

Before analyzing real data, we performed computer simulations to confirm whether the GLM could successfully decompose three different effects (AG, PO, and MG). We only considered the log-normal model for the simulations because both log-normal and negative binomial models assume additive effects of the three factors and their underlying assumptions are essentially the same. We evaluated a range of Cohen's $d$ ($1 \leq d \leq 5$), a ratio of the fixed effect to the standard deviation of statistical noise, for 2- and 5-times replicated datasets. In the GLM with Gaussian distribution, $p$ values monotonically decrease with $|d|$ and we expected that statistical power would increase with higher $d$ values and more replicates.

Our simulation using a duplicated dataset showed that we could accurately estimate each effect at $d = 5$ (Figure 2A), where the true positive rate of the effect was approximately 0.95 with a false discovery rate (FDR) of 0.05. As expected, the statistical power of the test increased remarkably with more replicates. We attained very high statistical power (true positive rate ~0.95) with five replicates when $d = 3$ (Figure 2B and Figure S2). We also tested whether any unbalanced effects could result in a biased estimation of each effect. Figure 2C shows results from five replicates wherein the $d$ values were 5 for AG and MG and 2 for PO. Despite the somewhat biased effect, we could accurately detect each significant effect.

*Analysis of Drosophila whole bodies*

We first analyzed two adult female *D. melanogaster* datasets, using a duplicated experimental design. In the log-normal GLM, after the initial filtering (see Materials and Methods), 6716 genes in the RAL799/RAL820 cross and 6971 genes in the RAL852/RAL324 cross were analyzed. We identified 776 and 1570 genes exhibiting signatures of the AG effect (FDR = 0.05) in the RAL799/RAL820 and RAL852/RAL324 crosses, respectively (Table 1). In the negative binomial GLM, 6536 genes in the RAL799/RAL820 cross and 6797 genes in the RAL852/RAL324 cross were analyzed. We identified 922 and 1732 genes exhibiting signatures of the AG effect (FDR = 0.05) in the RAL799/RAL820 and RAL852/RAL324 crosses, respectively (Table 2). Although none of the genes showed significant PO and MG effects with the log-normal GLM, 4–11 genes showed significant PO and MG effects with the negative binomial GLM.

Both methods agreed that 10%–20% of genes in the DGRP strains have a significant *cis*-regulatory effect. Among them, 221 and 400 genes showed significant AG effect in both reciprocal crosses in the log-normal and negative binomial GLM, respectively. On the other hand, none of the genes with significant PO and MG effects were overlapped between the reciprocal crosses. The results are provided in Supplementary Data 2–5.

*Analysis of mouse TSCs*

The second dataset was obtained in TSCs from mouse reciprocal cross Cast/B6 as reported by Calabrese et al., and composed three biological replicates. However, because one of replicate had been obtained in a previous study (CALABRESE *et al.* 2012), we only used the dataset with duplicates in our analysis. Using the log-normal GLM, we identified 1493, 273, and 4 genes with significant AG, PO, and MG effects, respectively (FDR <0.05), among 13,343 genes in this dataset.

Although the sexes of analyzed TSC samples are unknown, we expect that genes on the X chromosomes should show significant PO effect when the samples are males, because a male inherits the X chromosome only from a mother. Indeed, most of genes with significant PO effects (251/273) were located on the mouse X chromosomes, which implies that the samples were male TSCs. When we examined the pattern of gene expression on the Y chromosome and the expression level of *Xist* gene on the X chromosome, one of the TSC samples (GSM1561520) showed similar gene expression pattern to the male liver samples, further demonstrating that the TSC sample was from a male (data not shown). Therefore, we excluded the genes on the X chromosomes from further analysis. After the filtering, the number of genes with significant AG, PO, and MG effects were 1456, 22, and 4, respectively, in the log-normal GLM (Table 1).

Similar to the results of *Drosophila*, we observed slightly more genes with significant AG and PO effects using the negative binomial GLM after filtering out X chromosomal genes (Table 2); in total 2102 genes showed significant AG effect and 64 genes showed significant PO effect. However, the negative binomial GLM identified 393 genes with significant MG effects, considerably higher than those identified by the log-normal GLM. Detailed results are provided in Supplementary

Data 6 and 7.

*Analysis of mouse livers*

The third dataset comprised mouse liver expression data with six replicates, as performed by GONCALVES *et al.* (2012) using the same Cast/B6 reciprocal cross combination. Using the log-normal GLM, we identified 1608, 249, and 312 genes with significant AG, PO, and MG effects, respectively among the 12,293 genes in the liver dataset. Because the samples were derived from male livers, most of the PO genes were on the X chromosomes. After filtering out the genes on the X chromosome, the numbers of genes with significant AG, PO, and MG effects were 1584, 16, and 304, respectively (Table 1 and Supplementary Data 4). Likewise, among 11,169 autosomal genes, the negative binomial GLM identified 2014, 35, and 1355 genes with significant AG, PO, and MG effects, respectively (Table 2). Detailed results are provided in Supplementary Data 8 and 9.

*Evaluation of the log-normal and negative-binomial GLMs*

In both *Drosophila* and mice, the negative binomial model identified more genes with significant effects, which indicates that the negative binomial GLM has a lower rate of Type-II error and/or a higher rate of Type-I error. Although the difference is small for the AG and PO effects, the number of genes with significant MG effects considerably differs between the log-normal and negative binomial GLMs. Despite of some discrepancy, FDR-corrected *p*-values were highly correlated between the two models and the log-normal GLM gives more conservative estimate of *p*-values. Although the two methods have both advantages and disadvantages, we primarily show the results of negative binomial GLM in the following analyses.

*Comparison between mouse TSCs and livers*

Because the mouse TSC and liver data were obtained from the same reciprocal crosses, we contrasted the difference between the two tissues. In Figure 3, we present plots of the estimated effect sizes (fixed effect to the expression level in $\log_2$ scale) using the negative binomial GLM, for each gene in the TSCs and livers. We observed relatively small overlap (24%) of genes with significant AG effect between TSCs and livers, and many genes showed opposite AG effects in the two tissues. In contrast, although the significance level for the PO effect was different between the TSCs and livers, probably attributable to different sample size and noise level, the sign and size of effect were highly consistent between the two tissues. In contrast, a very small number of genes (29 genes) with significant MG effects were overlapped between the TSCs and livers, suggesting that the MG effect is highly tissue specific.

*Functional analysis of genes with significant effects*

We investigated whether there are significant enrichment of gene ontology (GO) terms among the genes with significant AG, PO, and MG effects. In *Drosophila*, none of the GO terms were overrepresented after controlling FDR = 0.05. In the mouse TSCs, only the genes with significant MG effects showed enrichment of annotated gene functions. Gene ontology terms neuron differentiation (GO:0030182) and neuron development (GO:404866) were slightly overrepresented in the genes with significant MG effects. In the liver, 30 Gene Ontology terms were significantly enriched among genes with the AG effect (FDR = 0.05); the most highly overrepresented gene category was oxidation reduction process (GO: 0055114). Most enrichment for the AG effect in the liver was related to oxygen metabolism process, protein binding activity, and membrane components (Table S2). Genes with significant PO effect in the liver did not exhibit any statistically significant enrichment. In contrast, genes with significant MG effects exhibited statistically significant enrichment gene annotation for 21 GO terms, mostly related to ribosomal and mitochondrial components (Table S3).

**DISCUSSION**

Here, we proposed a novel approach to decomposition of the three confounding effects affecting gene expression levels in reciprocally-crossed F1 hybrids. Although we applied the two different GLMs, log-normal and negative binomial GLMs, we first focused on the log-normal GLM and performed computer simulations, because the relationship between the effect size and error distribution in the log-normal model is much more intuitively understandable. Our simulation study showed that the efficiency of this method in the presence of sufficiently strong effects relative to statistical noises. In our duplicated *Drosophila* dataset, the average standard deviations of $\log_2$-transformed error were 0.255 for the RAL799/RAL820 reciprocal cross and 0.185 for the RAL852/RAL324 reciprocal cross. The higher error variance observed in the RAL799/RAL820 cross was likely responsible for the higher number of genes with significant AG effects in that line (Table 1 and Table 2). In mouse samples, the average standard deviations of $\log_2$-transformed error were 0.310 and 0.656 for TSCs and livers, respectively. As described above, we only focused on the log-normal model, but we should note that the assumptions for the distribution of biological and technical noises are different between the models, leading to the difference in statistical power.

Our analysis of two different reciprocal crosses of *Drosophila* largely corroborated Coolon et al.'s study that demonstrated an absence of genomic imprinting in *Drosophila* (COOLON *et al.* 2012). In addition, we did not find strong evidence of the MG effect in the adult female flies. However, the negative binomial GLM identified small number of genes with significant PO and MG effects. Because there was no overlap of those candidate genes between two different reciprocal crosses, we were not able to conclude whether those candidate genes were true positive genes. We also should note that studies to date have used only adult files. Therefore, further experiments based on samples from early developmental stages with more replicates are required to conclude the status of genomic imprinting and maternal effects in *Drosophila*.

In contrast to *Drosophila*, mouse datasets yielded several genes with significant PO and MG effects. Although the two datasets were conducted by different research groups, our comparison between TSCs and livers provided a good opportunity to investigate differences in each effect at the tissue and developmental-stage levels. Although many genes are imprinted in a tissue-specific manner (e.g., DECHIARA *et al.* 1991), our results showed a generally consistent genome-wide pattern of the PO effects across tissues and developmental stages (Figure 3). In contrast, small (24%) overlap between tissues was observed among genes with significant AG effects although similar number of genes were identified in both tissues. These results imply that a majority of *cis*-regulatory mutations are tissue specific. This pattern corroborates the modularity of gene regulation, wherein many mutations in *cis*-regulatory regions such as enhancers exhibit tissue-specific effects (WRAY 2007). Moreover, the number of genes with significant MG effect differed strikingly between TSCs and livers both in the log-normal and negative binomial GLMs. As we identified similar numbers of genes with significant AG and PO effects in both tissues, this difference might reflect important tissue-specific biological features. Although we cannot convincingly explain weaker MG effects in the TSC dataset, we suspect that TSCs, which are derived from embryos before implantation, spent less time for maternal-fetal crosstalk, compared to other fetal and adult tissues.

We examined weather our PO candidate genes in mice agree well with the 150 known imprinted genes (BLAKE *et al.* 2010). In TSCs, 26 out of 64 candidate genes were known as imprinted genes. *Ano1* and *Gab1* genes, which are not included in the list of 150 known imprinted genes but actually imprinted specifically in placenta (OKAE *et al.* 2012), were identified as PO-biased genes in the negative binomial GLM. Notably, the PO bias in *Ano1* was not detected by the study of Calabrese et al, which shared a part of the dataset with our study (CALABRESE *et al.* 2015). Likewise, in livers, 10 out of 35 candidate genes were known as imprinted genes. Among the 10 known imprinted genes, paternally-biased *Peg13* and maternally-biased *Rian* were not identified as imprinted genes in the study using the same dataset (GONCALVES *et al.* 2012). In addition to the known imprinted genes, we identified 60 candidate genes for the PO effect in TSCs and livers. Among them, *Gm11407* and *Snhg14* showed a signature of PO bias both in TSCs and livers. Although *Gm11407* is a pseudogene, *Snhg14* is a long-noncoding RNA located within an imprinted locus. Because human *SNHG14* is known to be imprinted (BABAK *et al.* 2008), mice *Shng14* is also likely imprinted. The list of imprinted genes identified in this study is shown in Table S4. Some known mouse imprinted genes did not achieve statistical significance probably because of our statistical method. For example, paternally imprinted *H19* genes did not meet our criteria for a significant PO effect. We examined expression data for this gene in TSCs and found that the imprinting status was not highly consistent among replicates. In addition, this gene was not expressed in livers. Therefore, our method requires sufficient replicates with good experimental conditions. In general, our method identified relatively fewer genes with PO effect than the previous studies using RNA-seq data (GREGG *et al.* 2010; WANG *et al.* 2011; DEVEALE *et al.* 2012; GONCALVES *et al.* 2012).

One of our most important methodologic achievements was the ability to evaluate the maternal effect (MG effect) without nuclear or embryo transplantation. Interestingly, the functional categories of genes with significant MG effects were very different between the TSCs and livers. In particular, genes with significant MG effects in livers contained many ribosomal and mitochondrial genes. Although it is reasonable that the mitochondrial genotype play an important role in the maternal effect, it is unlikely that the maternal cytosolic effect is still active in adult liver tissues. Therefore, the enrichment of ribosomal components in the genes with significant MG effects in livers should be a consequence of maternal effect during development. We note, however, that the maternal effect sizes were generally much smaller than those of the other two effects; even though statistical significance was detected, effect sizes of MG hardly exceeded 2, suggesting that the maternal effect is prevalent but has relatively minor effects on gene expression pattern (Figure 3C), compared with the AG and PO effects.

**CONCLUSION**

We have reported a novel method decomposing three confounding effects on allele-specific gene expression level in reciprocal crosses, and have demonstrated the effectiveness of this method using simulated data. Although data available is currently limited, this method yielded many biologically important observations in fruit flies and mice. This method will contribute greatly to our understanding of how genetic and epigenetic signals regulate patterns of gene expression and induce phenotypic diversity among tissues and individuals.

**ACKNOWLEDGMENTS**

We thank the Bloomington Stock Center for providing *Drosophila* stocks. Our thanks are due to Drs. Kenta Sumiyama and Hidemi Watanabe for valuable comments. This work was supported by a Grant-in-Aid for Scientific Research on Innovative Areas and Scientific Research (C) from the Ministry of Education, Culture, Sports, Science and Technology of Japan (KAKENHI Grant ID: 23113008 and 26440202).

**Figure 1**

Schematic representation of the generalized linear model (GLM) design. A hypothetical reciprocal cross between fly strains A and B is assumed. Black and white chromosomes represent the A and B genotype, respectively, and binary code specifies the allelic genotype (AG) effect (A: 0, B: 1). The parent-of-origin (PO) effect is set to 0 when the chromosome is inherited from the mother (left side of diploid chromosomes) and to 1 when the chromosome is inherited from the father (right side of diploid chromosomes). The maternal genotype (MG) effect is specified by the maternal genotype (A: 0, B: 1).

**Figure 2**

Methodologic evaluation using a simulated dataset. For each panel, the names heading rows represent effects given in the simulations; names heading columns represent effects estimated using the generalized linear model (GLM). Numbers in cells denote the fractions of correctly estimated effect among 1250 simulated genes. A) $d = 5$ with 2 replicates, B) $d = 3$ with 5 replicates, and C) $d = 5$ for allelic genotype (AG) and maternal genotype (MG) and $d = 2$ for parent-of-origin (PO) effects.

**Figure 3**

Comparison of effect sizes of mouse trophoblast stem cells (TSCs) and livers in the negative binomial GLM. The estimated effect size of each gene is indicated by a colored circle. The effect sizes of the TSCs and livers are shown on the $x$- and $y$-axes, respectively. Red circles represent genes with significant effects in both tissues and blue circles represent genes with significant effects in either tissue. Gens indicated by black circles did not exert significant effects. The allelic genotype (AG), parent-of-origin (PO), and maternal genotype (MG) effects are shown in panels A, B, and C, respectively.

**Table 1 Summary of the log-normal generalized linear model (GLM) analysis of *Drosophila* and mice**

| samples | *D. melanogaster* (female whole body) | | *Mus. musculus* (Cast/B6) | |
| --- | --- | --- | --- | --- |
| | RAL-799/RAL-820 | RAL-324/RAL-852 | TSC | liver |
| # of analyzed genes | 6176 | 6971 | 12,963 | 11,995 |
| AG (FDR = 0.05) | 776 | 1570 | 1456 | 1584 |
| PO (FDR = 0.05) | 0 | 0 | 22 | 16 |
| MG (FDR = 0.05) | 0 | 0 | 4 | 304 |

*AG, allelic genotype effects; PO, parent-of-origin effect; MG, maternal genotype effect

**Table 2 Summary of the negative binomial generalized linear model (GLM) analysis of *Drosophila* and mice**

| | *D. melanogaster* (female whole body) | | *Mus. musculus* (Cast/B6) | |
| --- | --- | --- | --- | --- |
| samples | RAL-799/RAL-820 | RAL-324/RAL-852 | TSC | liver |
| # of analyzed genes | 6536 | 6797 | 12,219 | 11,169 |
| AG (FDR = 0.05) | 922 | 1732 | 2102 | 2104 |
| PO (FDR = 0.05) | 5 | 15 | 64 | 35 |
| MG (FDR = 0.05) | 6 | 12 | 393 | 1355 |

*AG, allelic genotype effects; PO, parent-of-origin effect; MG, maternal genotype effect

$$♀×♂$$

A×B    B×A

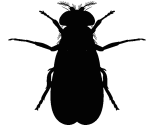      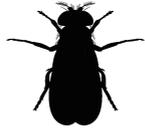

A B    B A

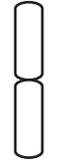 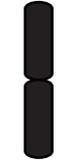     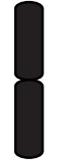 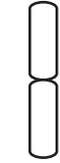

| | | | | | |
|---|---|---|---|---|---|
| AG | 0 | 1 | | 1 | 0 |
| PO | 0 | 1 | | 0 | 1 |
| MG | 0 | 0 | | 1 | 1 |

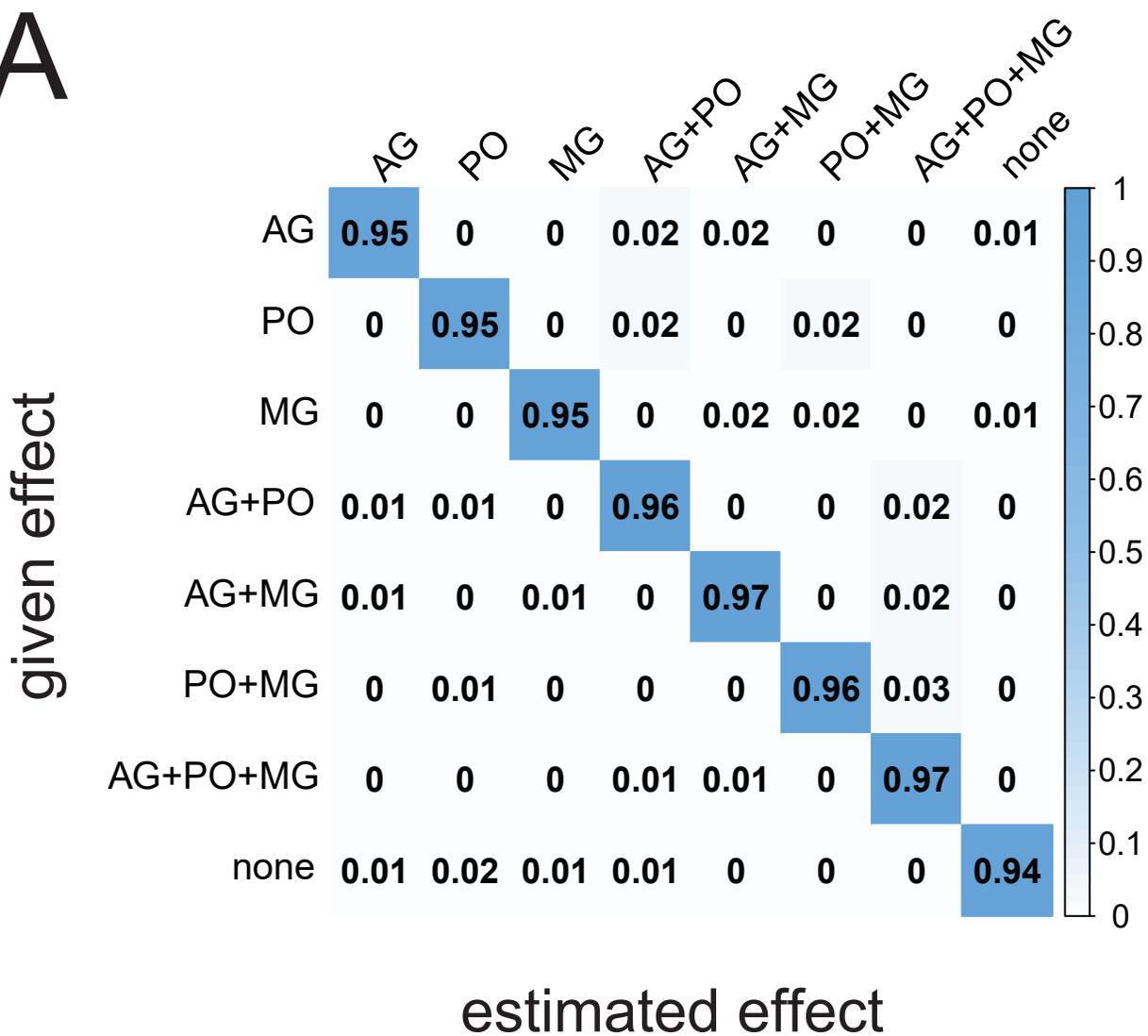

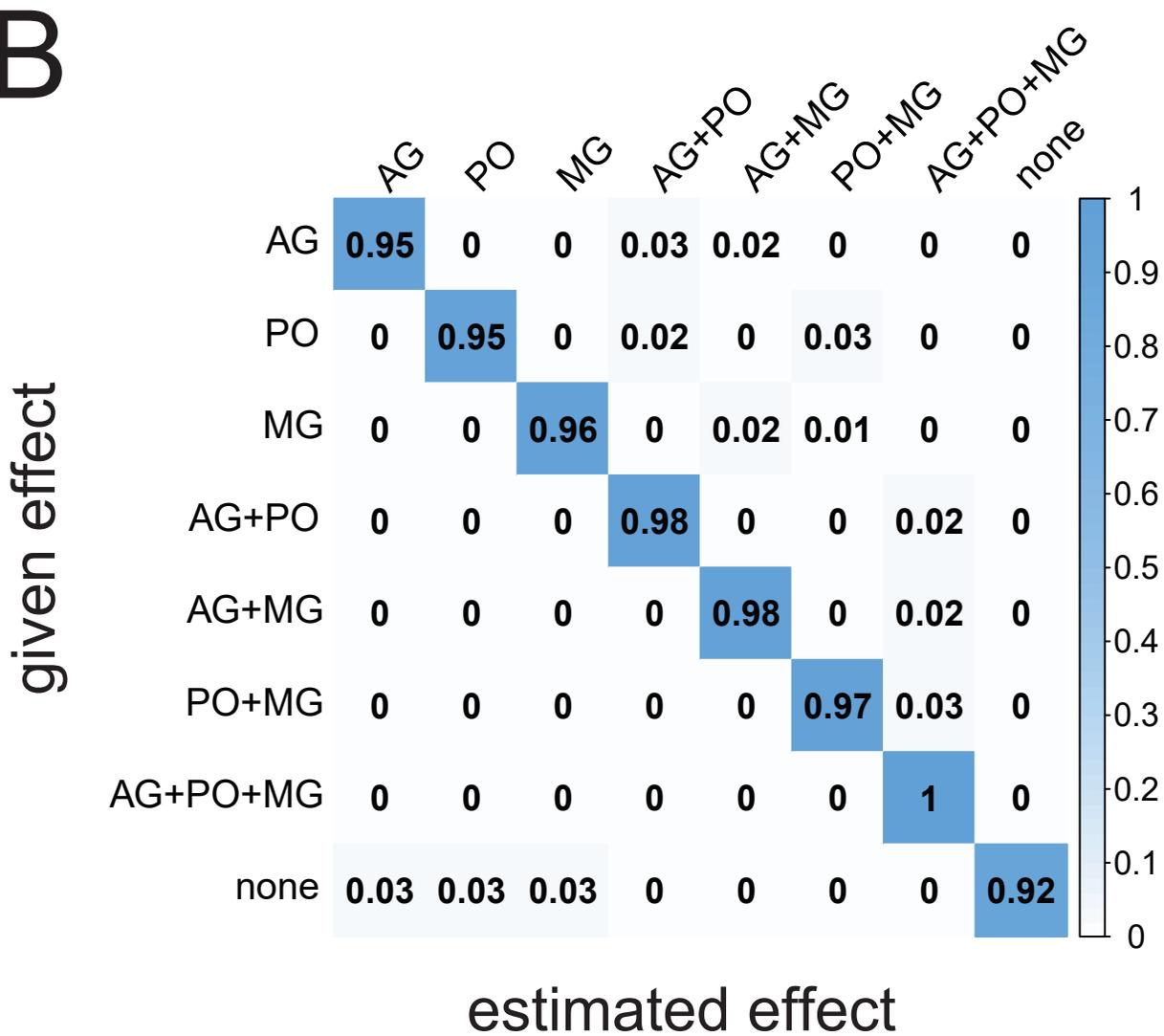

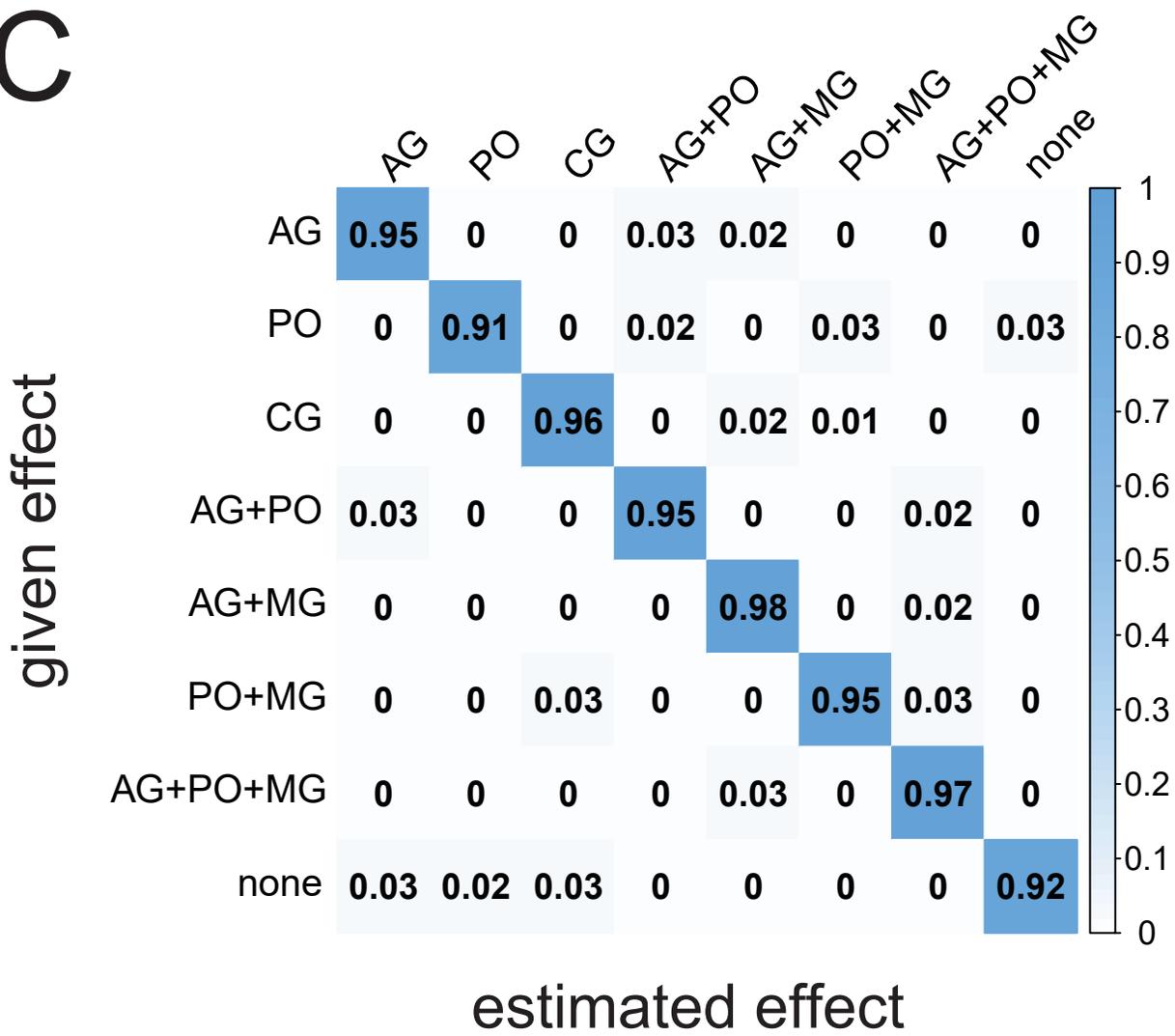

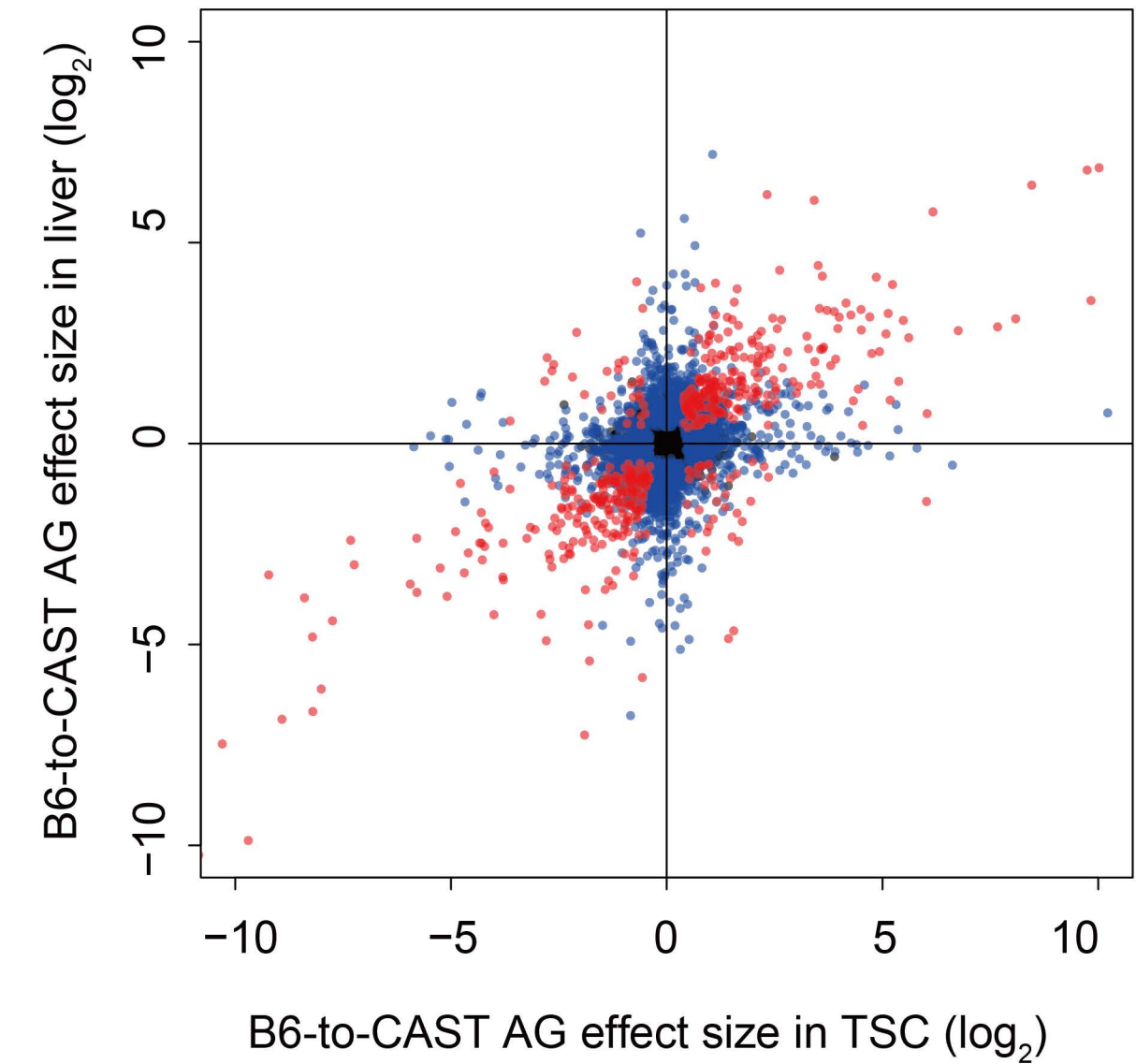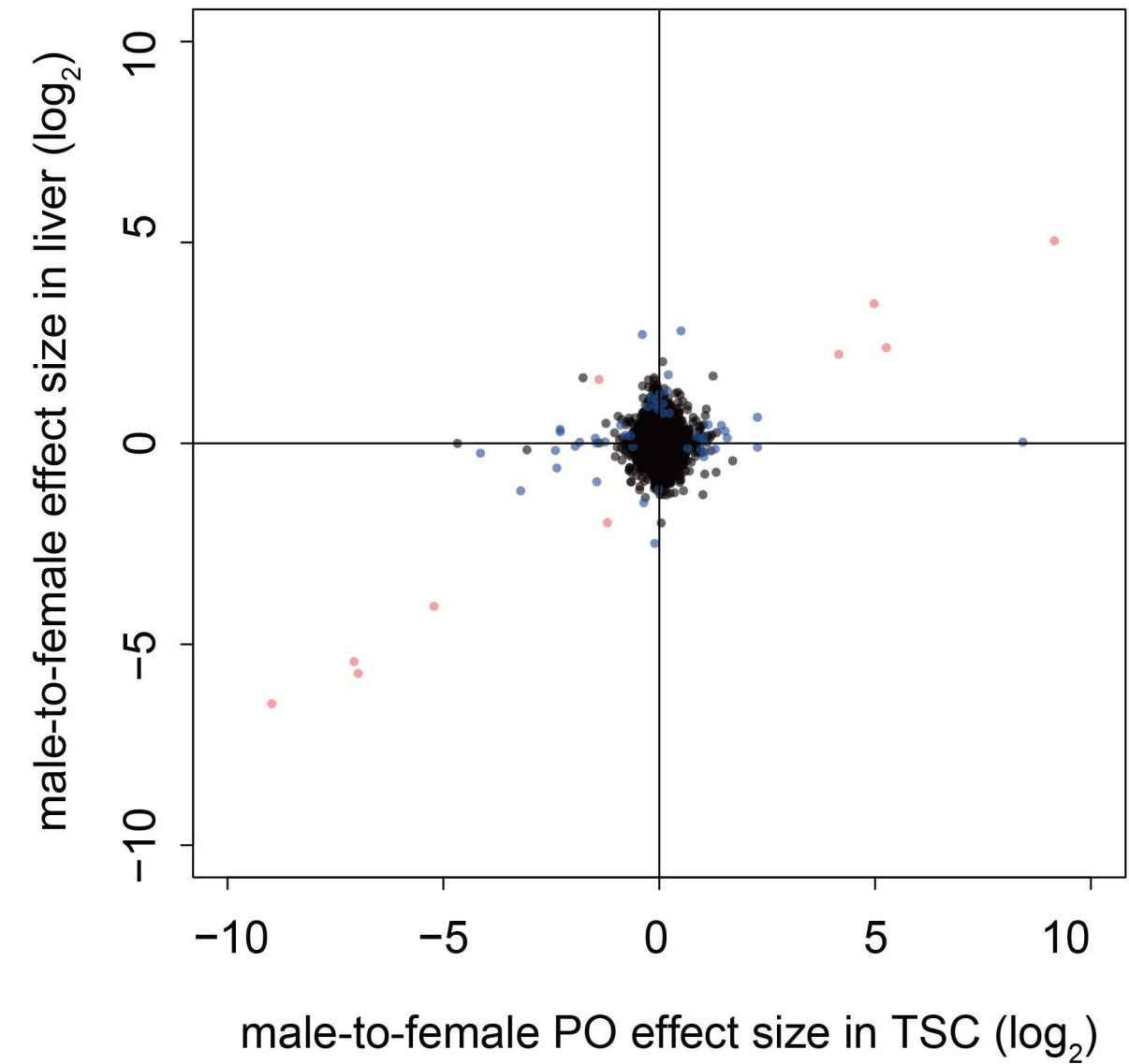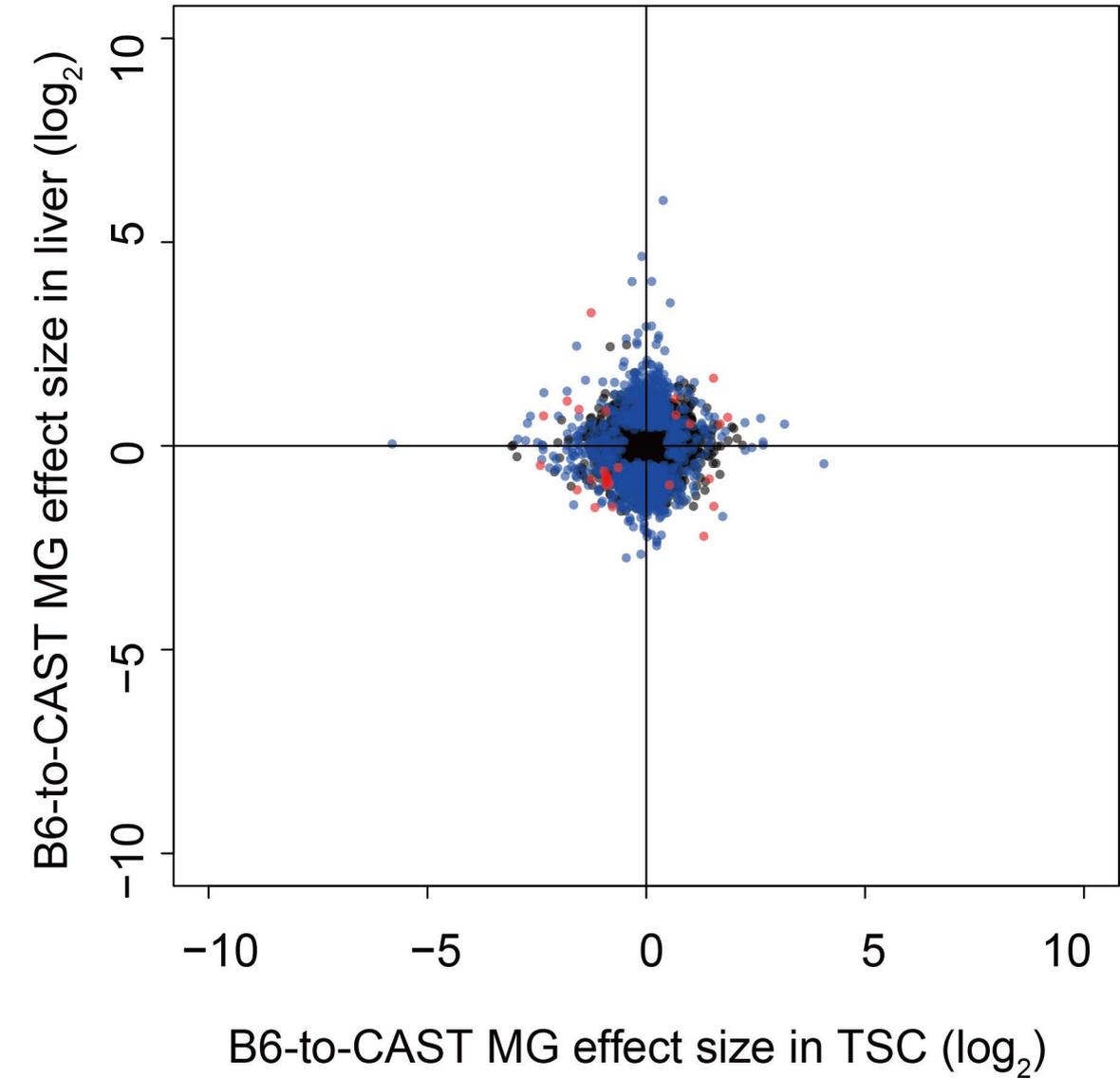

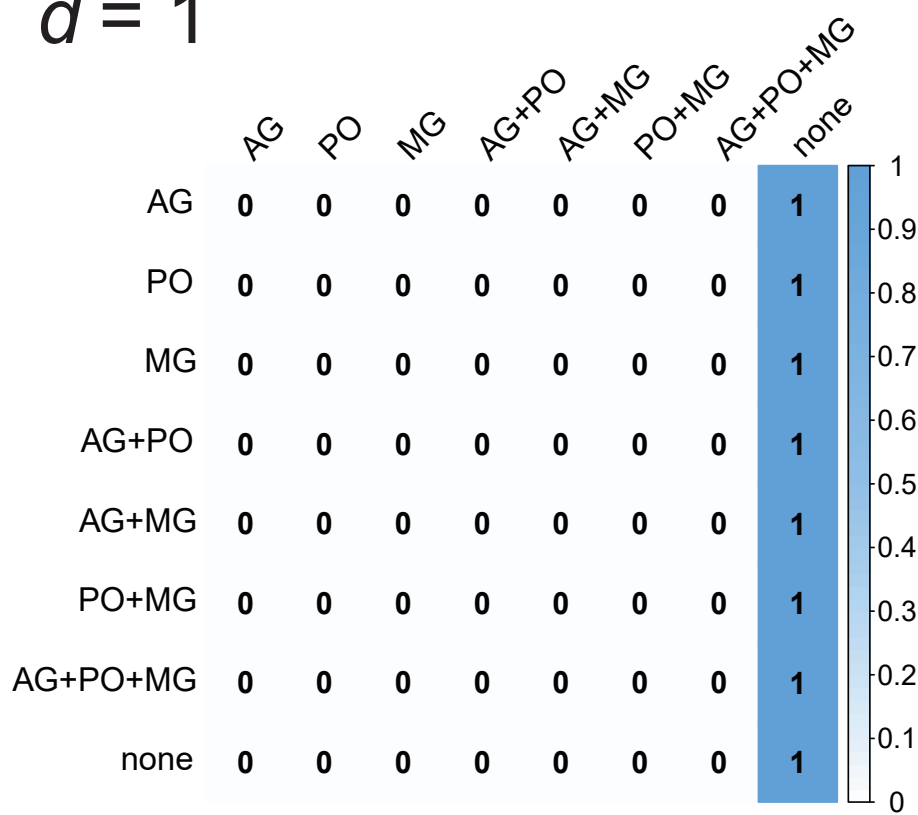

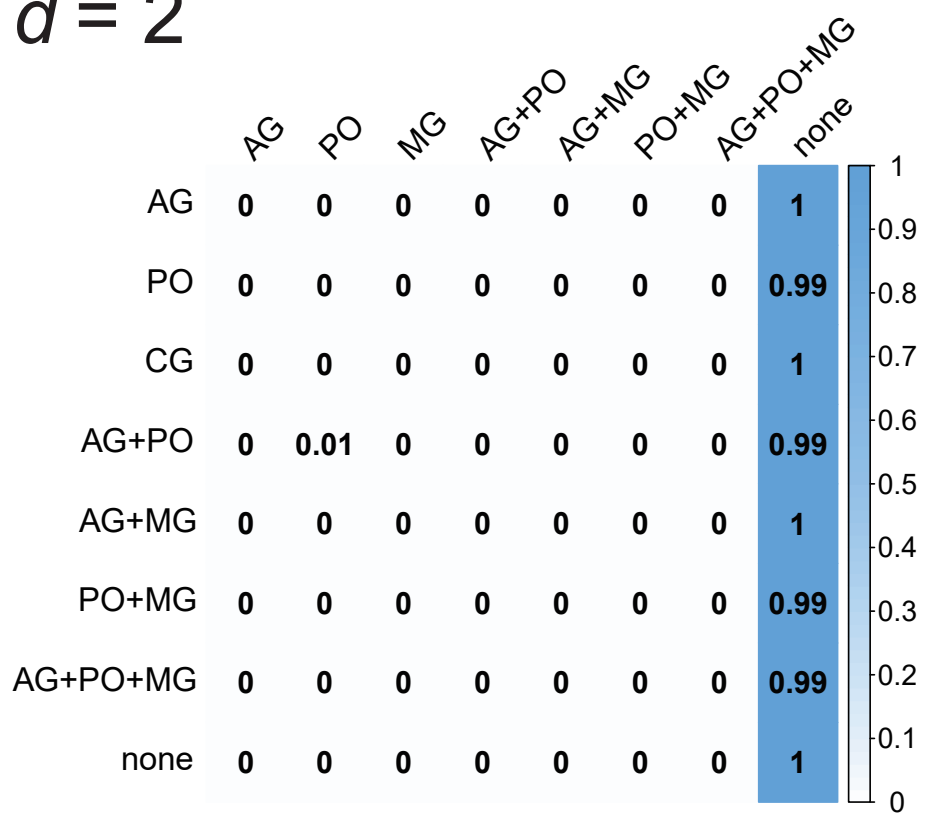

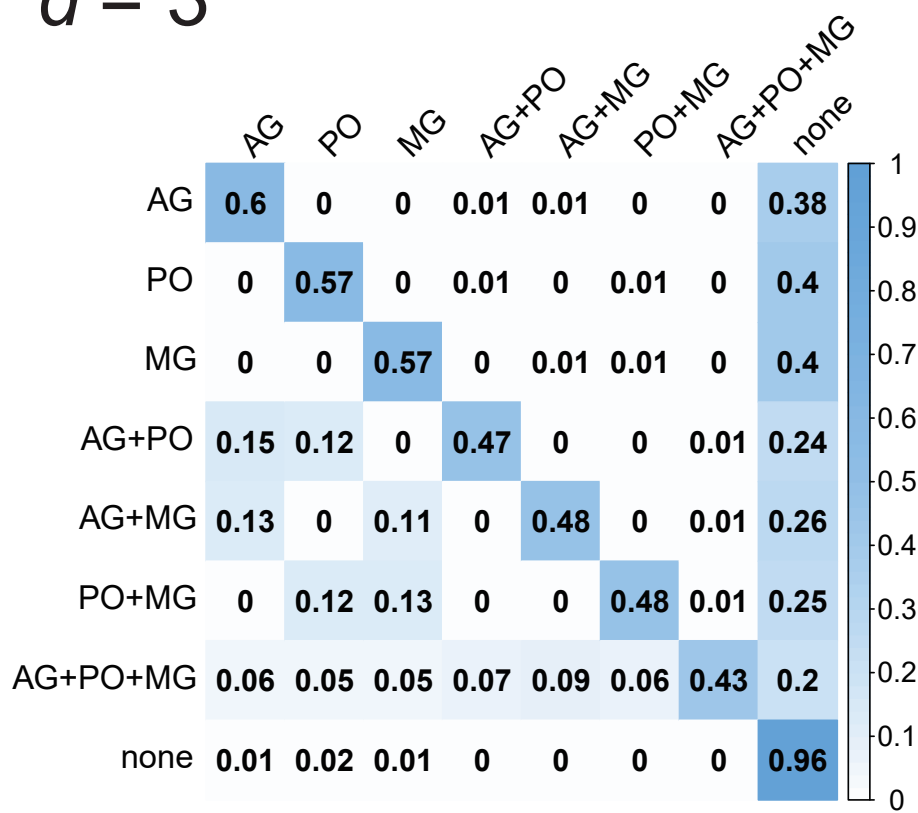

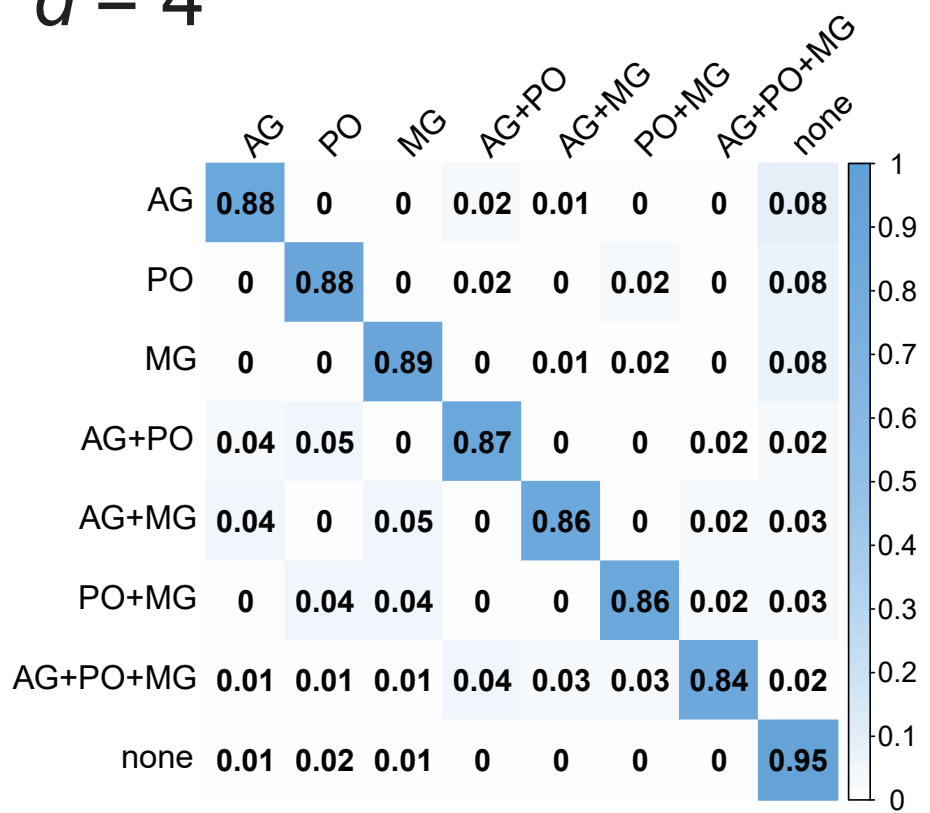

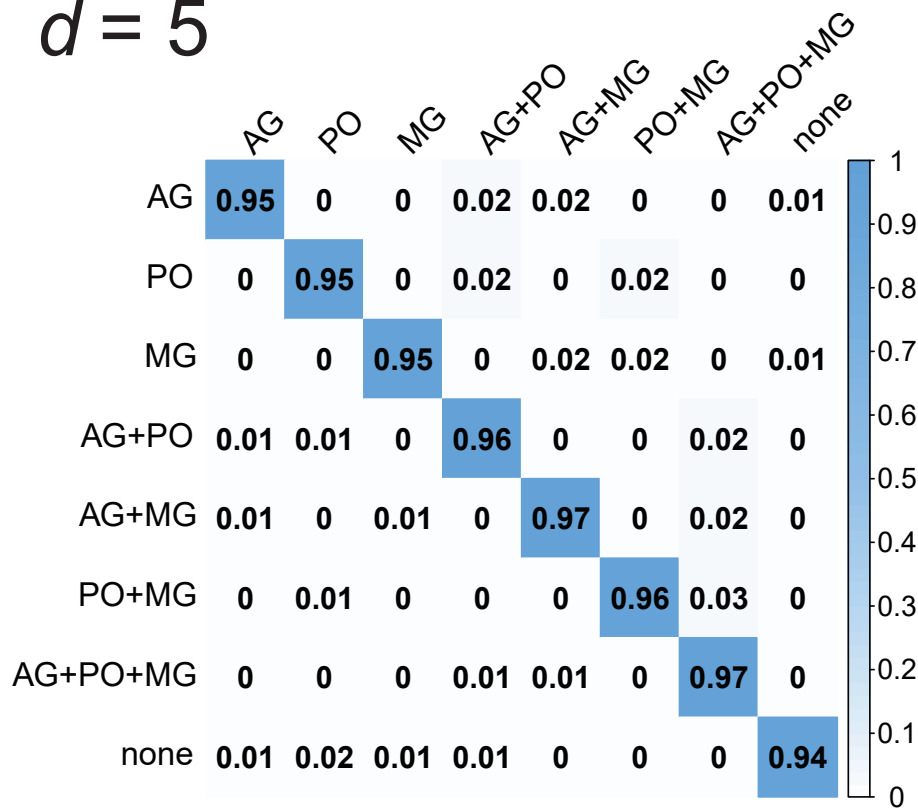

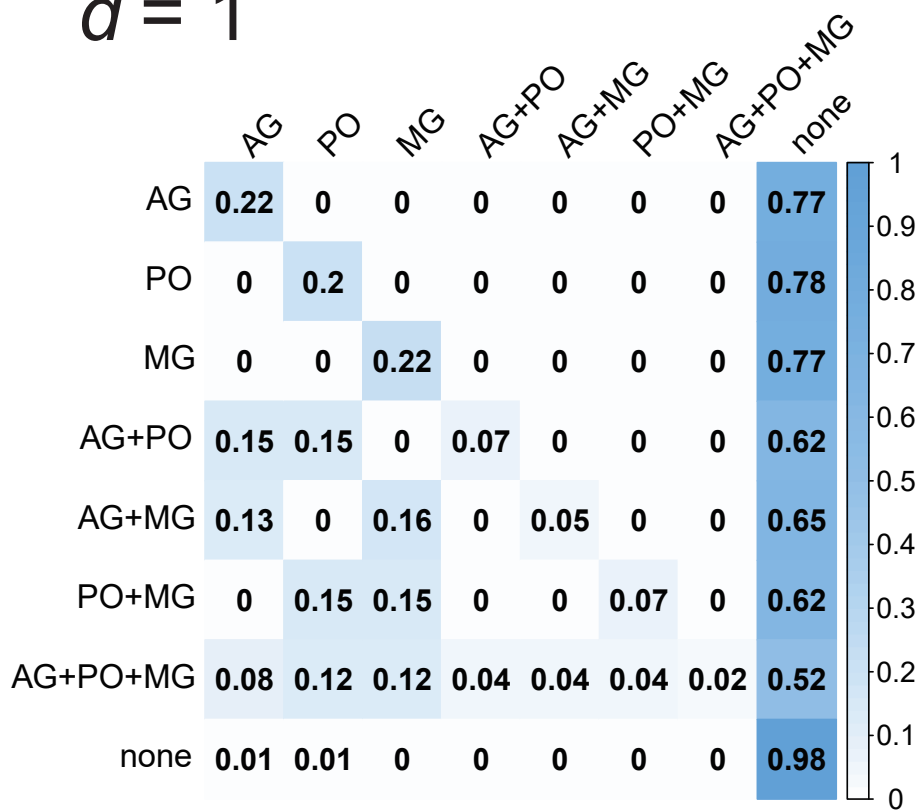

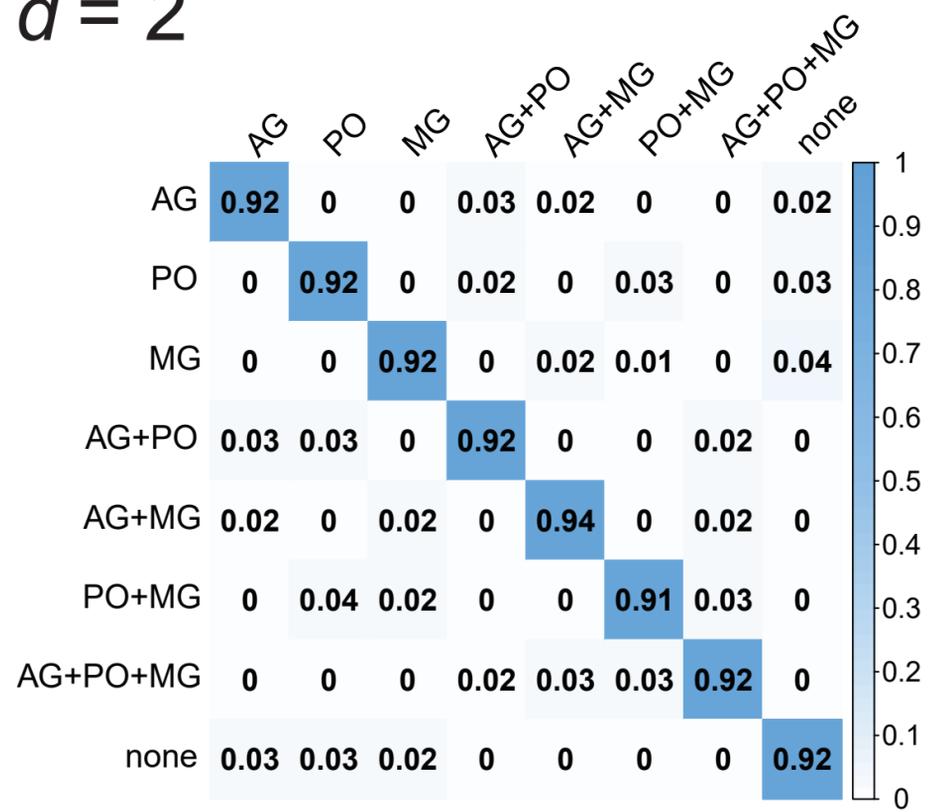

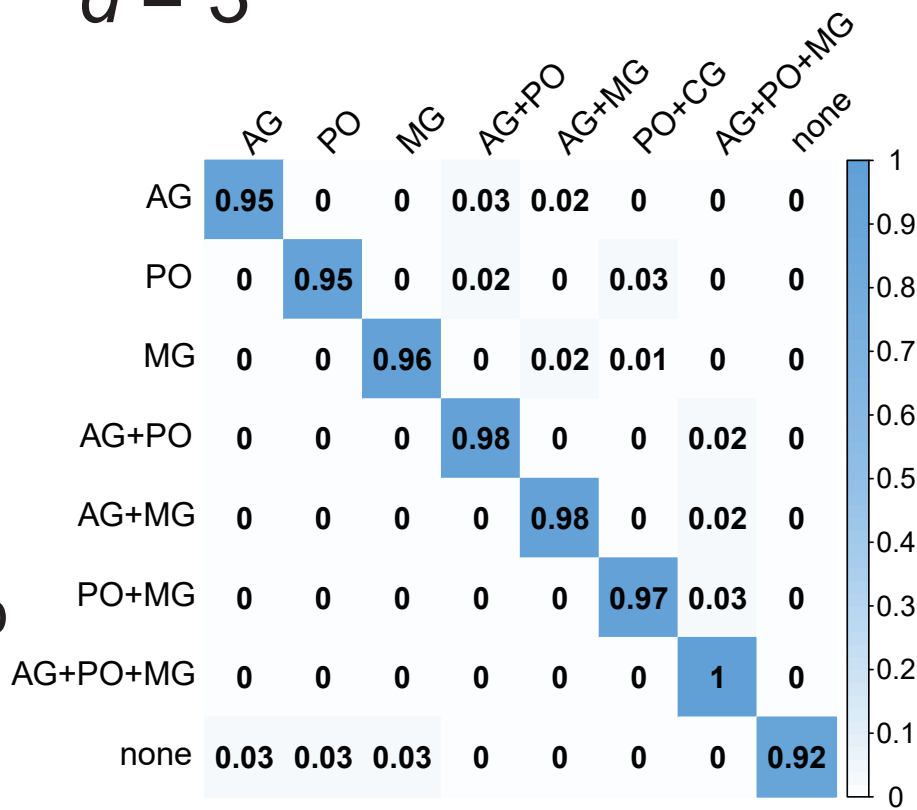

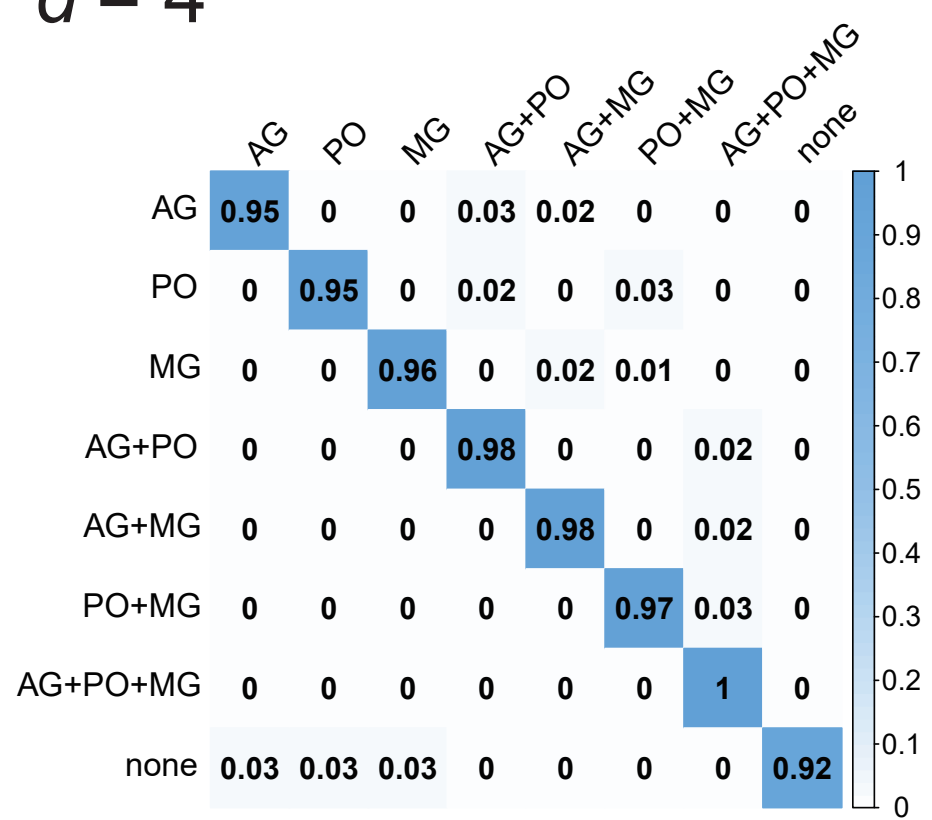

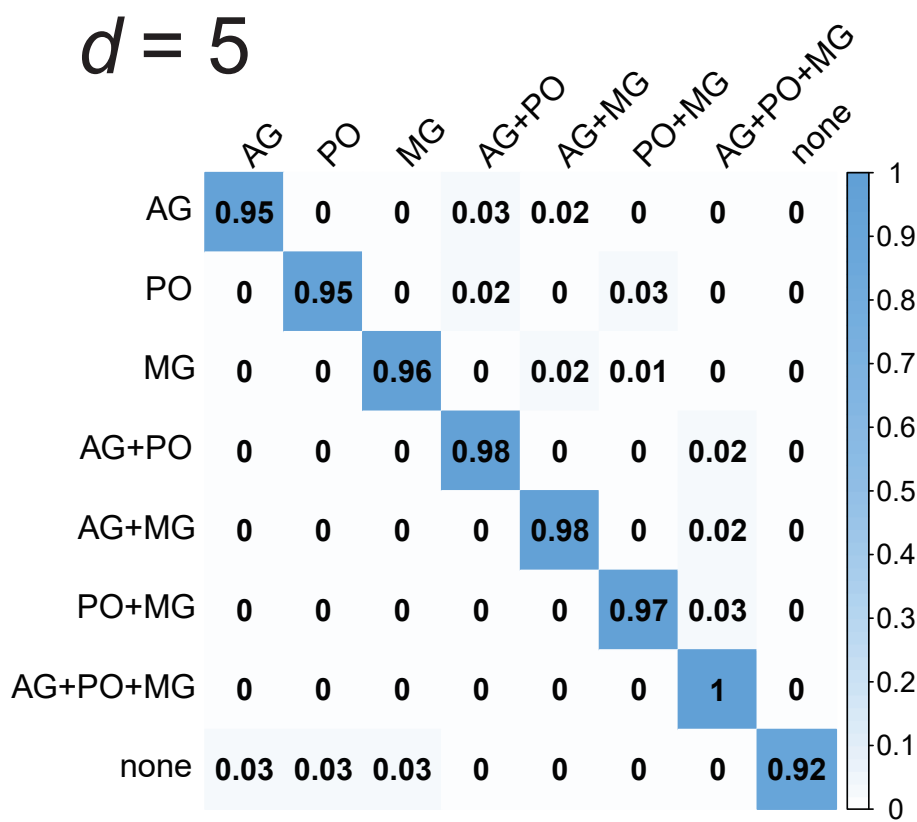

given effect

estimated effect

**Table S1. Accession numbers and Illuminabarcoding indices of libraries**

| Bioproject ID | sample name | sample accession | replication | DRA accession | Illumina barcode index |
|---|---|---|---|---|---|
| PRJDB5381 | RAL-799_RAL-820 | SAMD00069299 | 1 | DRX073117 | #14 |
| PRJDB5381 | RAL-799_RAL-820 | SAMD00069299 | 2 | DRX073125 | #4 |
| PRJDB5381 | RAL-820_RAL-799 | SAMD00069300 | 1 | DRX073118 | #16 |
| PRJDB5381 | RAL-820_RAL-799 | SAMD00069300 | 2 | DRX073126 | #6 |
| PRJDB5381 | RAL-324_RAL852 | SAMD00069301 | 1 | DRX073119 | #2 |
| PRJDB5381 | RAL-324_RAL852 | SAMD00069301 | 2 | DRX073127 | #6 |
| PRJDB5381 | RAL-852_RAL-324 | SAMD00069302 | 1 | DRX073120 | #12 |
| PRJDB5381 | RAL-852_RAL-324 | SAMD00069302 | 2 | DRX073128 | #5 |
| PRJDB5381 | RAL799 | SAMD00069303 | 1 | DRX073121 | #13 |
| PRJDB5381 | RAL799 | SAMD00069303 | 2 | DRX073129 | #2 |
| PRJDB5381 | RAL820 | SAMD00069304 | 1 | DRX073122 | #18 |
| PRJDB5381 | RAL820 | SAMD00069304 | 2 | DRX073130 | #7 |
| PRJDB5381 | RAL324 | SAMD00069305 | 1 | DRX073123 | #7 |
| PRJDB5381 | RAL324 | SAMD00069305 | 2 | DRX073131 | #4 |
| PRJDB5381 | RAL852 | SAMD00069306 | 1 | DRX073124 | #7 |
| PRJDB5381 | RAL852 | SAMD00069306 | 2 | DRX073132 | #4 |

**Table S2. Gene Ontology categories significantly enriched in the genes with AG effect in the mouse livers (negative binimial GLM).**

| GO Category | GO Term | # of genes | FDR-corrected *p* |
|---|---|---|---|
| Biological Process | GO:0055114~oxidation reduction | 135 | 7.31E-12 |
| Molecular Function | GO:0009055~electron carrier activity | 56 | 5.44E-10 |
| Celllar Component | GO:0005576~extracellular region | 154 | 8.33E-06 |
| Molecular Function | GO:0020037~heme binding | 34 | 1.13E-05 |
| Molecular Function | GO:0046906~tetrapyrrole binding | 34 | 4.70E-05 |
| Molecular Function | GO:0005506~iron ion binding | 63 | 2.23E-04 |
| Molecular Function | GO:0048037~cofactor binding | 53 | 3.17E-04 |
| Molecular Function | GO:0030414~peptidase inhibitor activity | 27 | 0.001538323 |
| Molecular Function | GO:0050660~FAD binding | 23 | 0.001978587 |
| Molecular Function | GO:0070330~aromatase activity | 14 | 0.002722371 |
| Molecular Function | GO:0004867~serine-type endopeptidase inhibitor activity | 21 | 0.003668986 |
| Biological Process | GO:0046394~carboxylic acid catabolic process | 27 | 0.003818393 |
| Biological Process | GO:0016054~organic acid catabolic process | 27 | 0.003818393 |
| Molecular Function | GO:0004364~glutathione transferase activity | 11 | 0.003852137 |
| Molecular Function | GO:0016712~oxidoreductase activity, acting on paired donors, with incorporation or reduction of molecular oxygen, reduced flavin or flavoprotein as one donor, and incorporation of one atom of oxygen | 15 | 0.004708334 |
| Molecular Function | GO:0004866~endopeptidase inhibitor activity | 25 | 0.00485135 |
| Biological Process | GO:0009063~cellular amino acid catabolic process | 20 | 0.0059055 |
| Biological Process | GO:0009310~amine catabolic process | 21 | 0.006245327 |
| Molecular Function | GO:0004857~enzyme inhibitor activity | 34 | 0.006778043 |
| Celllar Component | GO:0031224~intrinsic to membrane | 439 | 0.009249539 |
| Celllar Component | GO:0016021~integral to membrane | 419 | 0.009314458 |
| Celllar Component | GO:0005792~microsome | 40 | 0.01076923 |
| Celllar Component | GO:0042598~vesicular fraction | 41 | 0.012677847 |
| Celllar Component | GO:0005764~lysosome | 42 | 0.013216068 |
| Celllar Component | GO:0000323~lytic vacuole | 42 | 0.013216068 |
| Molecular Function | GO:0050662~coenzyme binding | 36 | 0.019653874 |
| Celllar Component | GO:0005773~vacuole | 43 | 0.028842934 |
| Celllar Component | GO:0044421~extracellular region part | 77 | 0.032827552 |
| Molcular Function | GO:0016229~steroid dehydrogenase activity | 10 | 0.03549423 |
| Celllar Component | GO:0000267~cell fraction | 88 | 0.037305212 |

**Table S3. Gene Ontology categories significantly enriched in the genes with MG effect in the mouse livers (negative binimial GLM).**

| GO Category | GO Term | # of genes | FDR-corrected $p$ |
|---|---|---|---|
| Molcular Function | GO:0003735~structural constituent of ribosome | 40 | 2.26E-08 |
| Celllar Component | GO:0005840~ribosome | 44 | 2.72E-06 |
| Celllar Component | GO:0070469~respiratory chain | 19 | 7.16E-06 |
| Molcular Function | GO:0005198~structural molecule activity | 51 | 4.76E-04 |
| Biological Process | GO:0006412~translation | 53 | 9.09E-04 |
| Celllar Component | GO:0044429~mitochondrial part | 78 | 0.004746971 |
| Celllar Component | GO:0019866~organelle inner membrane | 51 | 0.005158084 |
| Celllar Component | GO:0033279~ribosomal subunit | 18 | 0.005483124 |
| Celllar Component | GO:0005743~mitochondrial inner membrane | 48 | 0.008329068 |
| Celllar Component | GO:0005740~mitochondrial envelope | 58 | 0.008565734 |
| Celllar Component | GO:0031966~mitochondrial membrane | 55 | 0.008788277 |
| Celllar Component | GO:0015934~large ribosomal subunit | 12 | 0.010859552 |
| Celllar Component | GO:0022626~cytosolic ribosome | 8 | 0.021749824 |
| Celllar Component | GO:0030529~ribonucleoprotein complex | 66 | 0.022578141 |
| Biological Process | GO:0022900~electron transport chain | 24 | 0.023116349 |
| Biological Process | GO:0006091~generation of precursor metabolites and energy | 42 | 0.033047276 |
| Celllar Component | GO:0016469~proton-transporting two-sector ATPase complex | 12 | 0.034528878 |
| Celllar Component | GO:0032994~protein-lipid complex | 11 | 0.035227186 |
| Celllar Component | GO:0034358~plasma lipoprotein particle | 11 | 0.035227186 |
| Celllar Component | GO:0005739~mitochondrion | 156 | 0.036839455 |
| Celllar Component | GO:0045259~proton-transporting ATP synthase complex | 8 | 0.04649972 |

**Table S4. List of genes with significant PO effects in TSCs and livers (the negative binomial GLM)**

| gene | position in GRCm38 | logFC* TSC | logCPM† TSC | FDR‡ TSC | logFC liver | logCPM liver | FDR liver | known/unknown |
|---|---|---|---|---|---|---|---|---|
| *1700010I14Rik* | chr17:8988333-9008319 | -2.284 | 0.642 | 3.E-03 | NA | NA | NA | unknown |
| *1700028E10Rik* | chr5:151368675-151432118 | NA§ | NA | NA | 0.933 | 4.272 | 4.E-03 | uknown |
| *2410003L11Rik* | chr11:97598511-97622893 | 4.555 | 2.612 | 8.E-09 | NA | NA | NA | unknown |
| *A230059L01Rik* | chr1:146802965-146807340 | -1.462 | 3.545 | 9.E-05 | 0.006 | 2.621 | 1.E+00 | uknown |
| *A330032B11Rik* | chr19:37173843-37196541 | -0.358 | 2.781 | 1.E+00 | -1.477 | 2.934 | 4.E-02 | uknown |
| *Airn* | chr17:12741311-12860122 | 7.135 | 3.102 | 4.E-32 | NA | NA | NA | known |
| *Alg8* | chr7:97371606-97392185 | 0.094 | 6.242 | 1.E+00 | 0.988 | 3.089 | 5.E-02 | uknown |
| *Amigo1* | chr3:108186335-108192286 | -0.056 | 4.431 | 1.E+00 | 1.092 | 2.804 | 3.E-02 | uknown |
| *Ano1* | chr7:144588549-144751974 | -2.297 | 0.284 | 9.E-05 | 0.346 | 1.316 | 1.E+00 | uknown |
| *Atp6v0e2* | chr6:48537615-48541801 | -1.254 | 3.895 | 2.E-06 | 0.034 | 3.162 | 1.E+00 | unknown |
| *C1ra* | chr7:124512405-124523443 | 0.992 | 2.026 | 3.E-03 | 0.082 | 8.403 | 1.E+00 | uknown |
| *Ccdc114* | chr7:45924072-45948963 | -1.723 | 1.232 | 2.E-02 | NA | NA | NA | known |
| *Cd1d2* | chr3:86986551-86989780 | NA | NA | NA | 1.856 | 1.759 | 2.E-03 | uknown |
| *Cd81* | chr7:143052739-143067934 | -2.404 | 7.354 | 1.E-07 | -0.176 | 8.278 | 1.E+00 | uknown |
| *Cela1* | chr15:100674425-100687920 | NA | NA | NA | 0.966 | 5.060 | 2.E-03 | uknown |
| *Cenpk* | chr13:104228955-104249615 | 1.162 | 0.954 | 2.E-03 | NA | NA | NA | uknown |
| *Cideb* | chr14:55754045-55758458 | NA | NA | NA | 1.320 | 2.800 | 3.E-02 | uknown |
| *Cpsf4* | chr5:145167213-145182041 | 0.083 | 5.946 | 1.E+00 | 0.749 | 4.848 | 3.E-02 | uknown |
| *Ctsh* | chr9:90054152-90076089 | 0.865 | 4.215 | 4.E-03 | 0.170 | 7.936 | 1.E+00 | uknown |
| *Cyp4x1* | chr4:115106323-115134281 | -1.312 | 0.091 | 8.E-03 | NA | NA | NA | uknown |
| *Dact2* | chr17:14195231-14203831 | -0.779 | 5.032 | 4.E-02 | 0.202 | 4.801 | 1.E+00 | uknown |
| *Dlk1* | chr12:109452823-109463336 | -4.650 | 3.248 | 1.E-07 | NA | NA | NA | known |
| *E130012A19Rik* | chr11:97627389-97629702 | 0.658 | 6.721 | 7.E-03 | -0.119 | 2.165 | 1.E+00 | uknown |
| *Gab1* | chr8:80764438-80880519 | 2.276 | 8.602 | 3.E-23 | -0.097 | 3.670 | 1.E+00 | uknown |
| *Gm10499* | chr17:36141758-36145923 | 0.211 | 2.427 | 1.E+00 | 1.705 | 2.668 | 1.E-03 | uknown |
| *Gm11407* | chr4:80002331-80003388 | -1.395 | 9.797 | 2.E-08 | 1.589 | 10.861 | 1.E-03 | uknown |
| *Gm12763* | chr7:33653440-33659268 | NA | NA | NA | 3.893 | 0.590 | 2.E-05 | uknown |
| *Gm13247* | chr4:146502000-146539395 | 1.543 | 2.296 | 1.E-02 | NA | NA | NA | uknown |
| *Gm13261* | chr2:10339283-10374041 | 5.977 | 1.027 | 6.E-13 | NA | NA | NA | uknown |
| *Gm38893* | chr7:59974149-60005064 | 5.041 | 3.544 | 1.E-18 | NA | NA | NA | uknown |
| *Gm43841* | chr5:7276324-7276909 | NA | NA | NA | 3.413 | 3.380 | 3.E-07 | uknown |
| *Gng10* | chr4:59035088-59041903 | -1.481 | 3.848 | 2.E-02 | 0.126 | 2.851 | 1.E+00 | uknown |
| *Grb10* | chr11:11930508-12038683 | -3.209 | 9.946 | 5.E-37 | -1.182 | 1.996 | 1.E-01 | known |
| *H13* | chr2:152669461-152708670 | -1.201 | 7.391 | 1.E-11 | -1.973 | 7.764 | 1.E-23 | known |
| *Id1* | chr2:152736251-152737410 | 1.300 | 3.830 | 4.E-02 | -0.142 | 3.288 | 1.E+00 | uknown |
| *Igf2r* | chr17:12682406-12769664 | -8.977 | 10.002 | 1.E-83 | -6.478 | 5.827 | 2.E-64 | known |
| *Impact* | chr18:12972252-12992948 | 0.504 | 4.251 | 1.E+00 | 2.799 | 3.287 | 7.E-09 | known |
| *Itgam* | chr7:128062640-128118491 | NA | NA | NA | -1.574 | 0.895 | 1.E-02 | uknown |
| *Jade1* | chr3:41555731-41616864 | 1.575 | 8.317 | 1.E-02 | 0.132 | 5.144 | 1.E+00 | uknown |
| *Kcnq1ot1* | chr7:143212155-143296549 | 1.440 | 6.154 | 4.E-08 | 0.447 | 6.086 | 1.E+00 | known |
| *Lpar6* | chr14:73237895-73243294 | 1.057 | 2.812 | 1.E-02 | -0.204 | 4.213 | 1.E+00 | uknown |
| *Ly6g6c* | chr17:35065388-35070050 | -1.497 | 0.962 | 1.E-03 | NA | NA | NA | uknown |
| *Meg3* | chr12:109541001-109571726 | -6.974 | 7.131 | 2.E-09 | -5.726 | 5.263 | 1.E-32 | known |
| *Mest* | chr6:30723547-30748465 | 9.771 | 8.225 | 1.E-15 | NA | NA | NA | known |
| *Mrgpre* | chr7:143778363-143784500 | -1.945 | 0.026 | 2.E-07 | -0.071 | 4.112 | 1.E+00 | uknown |
| *Nfatc3* | chr8:106058840-106130537 | -0.896 | 6.822 | 5.E-03 | 0.453 | 4.865 | 1.E+00 | uknown |
| *Nmral1* | chr16:4710059-4719356 | -0.195 | 4.045 | 1.E+00 | 1.138 | 3.064 | 3.E-02 | uknown |
| *Pde10a* | chr17:8525372-8986648 | -2.240 | 5.202 | 6.E-09 | NA | NA | NA | uknown |
| *Peg10* | chr6:4747306-4760517 | 8.098 | 13.180 | 4.E-11 | NA | NA | NA | known |
| *Peg13* | chr15:72805600-72810324 | NA | NA | NA | 4.779 | 2.958 | 6.E-27 | known |
| *Peg3* | chr7:6703892-6730431 | 4.463 | 10.477 | 4.E-32 | NA | NA | NA | known |
| *Pik3cd* | chr4:149649168-149702571 | 0.190 | 4.835 | 1.E+00 | 1.305 | 1.836 | 5.E-02 | uknown |
| *Plagl1* | chr10:13060504-13131694 | 6.121 | 3.591 | 6.E-17 | NA | NA | NA | known |
| *Platr20* | chr11:51189833-51220418 | 3.500 | 1.775 | 3.E-05 | NA | NA | NA | uknown |
| *Platr4* | chr3:41484024-41493192 | 2.275 | 1.567 | 3.E-03 | 0.648 | 2.115 | 1.E+00 | uknown |
| *Ppip5k1* | chr2:121310561-121355396 | -0.262 | 4.789 | 1.E+00 | 0.911 | 3.909 | 5.E-02 | uknown |
| *R74862* | chr14:73021784-143053686 | -1.907 | 1.302 | 9.E-07 | NA | NA | NA | uknown |
| *Rasgef1b* | chr5:99217426-99729065 | -0.077 | 3.397 | 1.E+00 | 0.951 | 3.284 | 3.E-02 | uknown |
| *Rbm15* | chr3:107325421-107333673 | -0.034 | 6.253 | 1.E+00 | 0.973 | 3.632 | 4.E-02 | uknown |
| *Rian* | chr12:109603940-109661716 | -7.068 | 7.494 | 7.E-08 | -5.432 | 3.976 | 1.E-16 | known |
| *Rnf2* | chr1:151458004-151500955 | 0.004 | 6.581 | 1.E+00 | 1.231 | 3.310 | 1.E-02 | uknown |
| *Rpl37* | chr15:5116613-5119140 | 1.035 | 6.684 | 2.E-03 | -0.332 | 6.444 | 1.E+00 | uknown |